\documentclass[12pt,a4paper]{article}
\usepackage{color}
\usepackage{graphicx}
\usepackage{bm}
\newcommand{\be}{\begin{eqnarray}}
\newcommand{\ee}{\end{eqnarray}}

\newcommand{\rar}{\rightarrow}
\newcommand{\bi}{\bibitem}
\newcommand{\lrar}{\leftrightarrow}

\newcommand{\mnu}{m_\nu}

\newcommand{\dm}{\delta m^2}

\definecolor{gold}{rgb}{0.89,0.78,0}
\definecolor{grn05}{rgb}{0,0.5,0}

\begin{document}

{
\title{{
{
Cosmology and Neutrino Properties
}}}
}
\author{
{A.D. Dolgov}
\\[5mm]
{
\it ITEP, 117218, Moscow, Russia}\\
{\it University of Ferrara and INFN, FE 40100, Italy}
}

\maketitle

\begin{abstract}

This is a brief review for particle physicists on cosmological impact of neutrinos 
and on restrictions on neutrino properties from cosmology. The paper includes  
discussion of upper bounds on neutrino mass and possible ways to relax them, methods 
to observe the cosmic neutrino background, bounds on the cosmological lepton asymmetry 
which are strongly improved by neutrino oscillations, cosmological effects of breaking 
of spin-statistics theorem for neutrinos, bounds on mixing parameters of active and
possible sterile neutrinos with the account of active neutrino oscillations, bounds 
on right-handed currents and neutrino magnetic moments, and some more.

\end{abstract}

\section{Introduction \label{s-intro}}

Neutrino is the weakest of all known elementary particles,
but despite that, or maybe because of that, cosmological impact
of neutrinos is significant. Neutrinos are important for cosmology
and, vice versa, astronomy allows to measure neutrino properties
with precision which is in many cases higher than the precision of
direct experiments. 

{{
Neutrinos are the second most abundant particle in the universe
(after photons of CMBR), their total number density, with
antineutrino included, is
\be
{{ \Sigma_j\, (n_\nu+n_{\bar\nu})} {\approx 340 /{\rm  cm}^3}},
\label{sigma-n-nu}
\ee
almost equally shared between all three active neutrino 
species ${{ \nu_e}}$, ${{ \nu_{\tau}}}$, and  $\nu_{\tau}$. 
For comparison the average cosmological number density of the normal 
baryonic matter is
\be
{ n_B \approx 2.5\times 10^{-7} /{\rm cm}^3}.  
\label{n-B}
\ee
The average momentum of cosmic neutrinos is very low, about 
$6.1^oK \approx 5.3\cdot 10^{-4} $ eV.
The cosmic neutrino background (C${\nu}$B) is not observable 
directly at the present time, because of the very weak interactions of 
these low energy neutrinos. 
Still their cosmological impact is profound.

In this talk I will consider cosmological effects produced by neutrinos 
and cosmological bounds on neutrino properties. More detailed
discussion can be found in the review paper~\cite{nu-rev}, but
presentation here contains some recent development.

The content of the talk is the following: \\
{1. Cosmological bounds on neutrino masses, ${m_\nu}$. }\\
{2. Possibility of direct registration of C$ \nu$B.}\\
{3. Restriction on the number of neutrino species from BBN and CMBR.}\\
{4. Neutrino statistics and cosmology.}\\
{5. Neutrino oscillations, BBN, and cosmological lepton asymmetry.}\\
{6. Right handed currents and magnetic moment of neutrinos.}\\

}}

\section{Thermal history of neutrinos and Gerstein-Zeldovich
bound \label{gz}}

In the early universe at $T>1 $ MeV neutrinos were in
thermal equilibrium with electron-positron pairs and photons. 
Correspondingly their number density was
\be{
n_\nu = (3/8) n_\gamma}
\label{n-nu-0}
\ee
for each {left-handed} neutrino flavor, 
${\nu_e,\nu_\mu,\nu_\tau}$,
under assumption of vanishing asymmetry between neutrinos 
and antineutrinos, i.e. 
\be
{{ n_\nu = n_{\bar\nu}}}.
\label{n-nu-n-nubar}
\ee
The condition for thermal equilibrium is found by comparing the
neutrino reaction rate, {${{ \Gamma = \sigma n \sim G^2_F T^5}}$},
with the cosmological expansion rate, $H = \dot a/a \sim T^2 /m_{Pl}$.
If {${ \Gamma > H}$}, thermal equilibrium
is established. This took place at $T\geq 1$ MeV.

More accurate treatment based on solution of the kinetic equation
governing neutrino distribution~\cite{nu-rev} shows that $\nu_e$
decoupled from the electron-positron pairs at $T\approx 2$ MeV,
while $\nu_{\mu,\tau}$ decoupled at $T \approx 3$ MeV. Interactions
between neutrinos maintain their kinetic equilibrium down to lower
temperatures, $T_{\nu_e} = 1.3$ MeV and $T_{\nu_\nu,\nu_\tau} = 1.5$
MeV for $\nu_e$ and $\nu_{\mu,\tau}$ respectively. Below these temperatures
neutrinos practically decoupled from the plasma and free streamed with
the speed of light till they became nonrelativistic. 

At smaller temperatures, ${T\leq m_e}$, photons were heated by 
${{ e^+e^-}}$-annihilation, while ${\nu}$ were not, because they
had already decoupled from the electromagnetic component of the 
primordial plasma. The increased number density of photons had
led to the decrease of the neutrino-to-photon ratio:
\be
n_\nu + n_{\bar\nu} = (3/11) n_\gamma = 112 /{\rm  { cm^3}}
\label{nu-to-gamma}
\ee
The coefficient $3/11$ is obtained from the entropy conservation
and the numerical value of the neutrino+antineutrino number density 
is presented for the present day universe and obtained from the
measured number density of photons in cosmic microwave background
radiation (CMBR), $n_\gamma = 410.5\pm 0.5/$cm$^3$. 

As a result of the photon heating the temperature of neutrinos
dropped down with respect to the photon temperature:
\be
T_\nu = 0.714 T_\gamma = 1.945 K^o = 1.68\cdot 10^{-4}\,{\rm eV},
\label{T-nu}
\ee
where the second equality is written for the present day neutrino
temperature calculated from the known value of $T_{gamma} = 2.725 K^0$. 
One should remember, however, that for massive neutrinos the distribution
function has the form:
\be
f_\nu = \left[ \exp (p/T) + 1 \right]^{-1}.
\label{f-nu}
\ee
It is not equilibrium distribution because in the equilibrium one
there should be the neutrino energy, $E$, instead of momentum, $p$, as in
eq.~(\ref{f-nu}). So, strictly speaking, parameter $T$ here is not 
temperature. The difference, however, is essential only at low
temperatures, $T\sim m_\nu$. 

Knowing the number density of neutrinos in the present day universe
we can calculate their energy density and from the condition that
the latter does not exceed the measured energy density of matter we 
obtain the cosmological bound on neutrino mass: 
\be
\Sigma_a m_{\nu_a} < 94 \Omega_m h^2\,\,{\rm eV}
\label{GZ}
\ee
where $h = H/100 km/s/Mpc$ is the dimensionless Hubble parameter
and $\Omega_m$ is the fraction of the energy of matter relative
the critical energy density, $\rho_c = 3H^2 m_{Pl}^2/8\pi$.
This bound was derived by Gerstein ans Zeldovich in 1967~\cite{gz}.

Using the observational data: ${ \Omega_m \leq 0.25}$ and ${ h = 0.7}$,
we find:
\be
\Sigma_a m_{\nu_a} < 11.5 \,\, {\rm eV}
\label{sigma-m-nu1}
\ee
For almost equal masses of neutrinos (as we know from neutrino oscillations):
\be
 m_{\nu_a} < 3.9\,{ {\rm {eV}}}
\label{m-nu-a}
\ee
This bound can be further improved by factor $\sim$3 because
too large energy density of neutrinos would inhibit large scale
structure (LSS) formation at relatively small scales and at hight 
redshifts, $z\geq 1$. In simple words, neutrinos are fast and reluctant 
to form gravitationally bound systems.
To eliminate such an undesirable property 
the mass density of hot dark matter, created by cosmic neutrinos should
not exceed $\sim$30\% of the total matter density. Correspondingly:
\be
m_{\nu_a} < 1.2\,\,  {\rm eV}
\label{m-nu-hdm}
\ee
This is a robust and quite strong bound, stronger than that obtained in the
direct experiments. 

The best bounds on neutrino mass, obtained from 
tritium decay experiments (Troitsk and Meinz) is~\cite{pdg}:
\be
m_{\nu_e} < (2-3)\,\, {\rm { eV}}
\label{m-nu-e}
\ee
Though this bound is usually presented as a bound on the mass of electronic
neutrino, it is inaccurate because, as we know from the measured neutrino
oscillations, mass eigenstates are
strongly different from the flavor eigenstates, i.e. from $\nu_{e,\mu,\tau}$.
Neutrino oscillations allow to measure only two mass differences of three
mass eigenstates, but not the absolute value of the mass. 
Neutrino oscillation data are best fitted with~\cite{pdg}:
\be {{
\delta m^2_{solar} = (5.4 - 9.5) \cdot 10^{-5}}\,\, {\rm { {eV}}^{ 2}}}
\nonumber \\
{ \delta m^2_{atm} = (1.2 - 4.8) \cdot 10^{-3}}\,\, {\rm { eV}}^{ 2}
\label{delta-m}
\ee
Presently telescopes allow to weight neutrinos more accurately than
direct experiments.
Astronomy can be sensitive to 
\be
m_\nu \sim { \rm{(a\,\,\, few)}}{ \times 0.1}\, {\rm eV}, 
\label{m-nu-cosm}
\ee
based on combined data on {LSS} and {CMBR} (see sec.~\ref{s-mass-cncl}).

{\it {Historical remark:}} the paper by Gerstein and Zeldovich
``Rest mass of muonic neutrino and cosmology'' was published
in 1966~\cite{gz}. Six years later a similar paper by 
Cowsik and McClelland~\cite{cow} ``An upper limit on the neutrino rest
mass'' was published. In many subsequent works the cosmological bound
on neutrino mass is quoted as ``Cowsik-McClelland bound''. This is not just,
however, firstly, because the GZ paper was much earlier and, secondly, in the
paper by Cowsik and McClelland the effect of photon heating by $e^+e^-$-annihilation
was disregarded and both helicity states of neutrinos, left-handed and right-handed,
were assumed to be equally populated. This incorrectly gives rise to 7 times stronger
bound. 

We know that only left-handed neutrinos participate in weak interactions.
If neutrino mass is nonzero the other, right-handed state, must exist. However,
right-handed neutrinos could be in equilibrium only at very high temperatures
and even if the were abundantly created at some early cosmological stage,
they would be strongly diluted by entropy released in massive particle
annihilation and so their number density would normally be negligible
at $T\sim $ MeV.

\subsection{Is it possible to relax the GZ bound? \label{ss-relax}}

Let us critically discuss essential assumptions used in the derivation of
the GZ bound and check if they can be modified in such a way that
the present day number density of C$\nu$B would be noticeably smaller
than the standard one (\ref{nu-to-gamma}) and thus would 
allow for a larger neutrino mass.

The obtained upper bound on $m_\nu$ is based on the following:\\
1.  {\it Thermal equilibrium in the early universe.} 
It is surely true in the standard cosmology. Neutrinos would not
be produced in equilibrium amount if the primeval plasma temperature 
were never larger than a few MeV~\cite{kolb}. 
It could be realized in inflationary models 
with anomalously low (re)heating temperature. 
However, a smaller number
density of neutrinos during big bang nucleosynthesis (BBN) would distort
successful predictions for light element abundances. The problems with 
baryogenesis would be also serious.\\
2. {\it Non-vanishing lepton asymmetry.} We assumed that number density
of neutrinos is equal to that of antineutrinos. If the primeval plasma has a
non-zero lepton asymmetry this equality would be broken. However, in thermal
equilibrium the total number density $n_\nu+n_{\bar \nu}$ would be larger in
asymmetric case and the upper bound on $m_\nu$ would be stronger. Moreover, 
as we see below, the cosmological lepton asymmetry is strongly bounded from
above and cannot noticeably change the neutrino number density.\\
3. {\it Conservation of the ratio $n_\nu/n_\gamma$ from 1 MeV to the present day.}
If there are some extra sources of heating of the photon plasma below
MeV, the ratio ${{ n_\nu/n_\gamma}}$ would be smaller than (\ref{nu-to-gamma}).
This can be achieved e.g. by electromagnetic decays of new light long-lived particles.
Care should be taken of BBN and the frequency and angular spectra of CMBR. In
particular, the ratio $n_B/n_\gamma$ measured at BBN and CMBR cannot differ more
than by factor 2. It seems to be difficult and one has to invoke some more new
physics, but maybe this is not excluded.\\
4. {\it Neutrino stability at the cosmological time scale.}
It is assumed that neutrinos created in the plasma at $T\sim 1$ MeV
survived up to the present time, i.e. ${ \tau_\nu \geq t_u \sim 10^{10}}$ years. 
Bearing in mind a small neutrino mass (\ref{m-nu-e}), it is difficult to avoid 
the conclusion about a long life time of $\nu$. To facilitate neutrino decay
new interactions are necessary. Even if such exotics exists, it does not lead
to relaxation of GZ limit if neutrino decays into a very light or massless
particle (e.g. photon) plus another a little lighter neutrino, because in this
case the total number of neutrinos is conserved. As we see from 
eq.~(\ref{delta-m}) the mass difference is much smaller than eV and thus
the upper bound on the neutrino mass remains the same. However, if the
decay goes into a new sterile neutrino, the cosmological bound on the masses
of normal active neutrinos could be weaker.\\
5. {\it An absence of new strong interactions between neutrinos at low energies.}
A new interaction may enhance $\bar\nu \nu$-annihilation into new very light
particles, so the neutrino number density today would drop down. To this end
a very light or massless boson is needed. Such bosons are practically excluded by
stellar cooling, see e.g.~\cite{farzan}. \\
6. {\it Absence of right-handed neutrinos.} If both helicity states of neutrinos
would be populated the bound would be stronger.\\
7. Presented below stronger bounds, based on angular spectrum of CMBR and analysis
of LSS, depend upon the values of the cosmological parameters, on the spectrum
of primordial perturbations, and on the type of dark matter.
A nonstandard spectrum of density perturbations at small scales
may change these strong bounds on $m_\nu$ but would not
influence the classical GZ bound.

\subsection{ Non-equilibrium cosmic neutrinos \label{ss-noneq-nu}}

Usually the cosmological expansion does not destroy equilibrium 
distribution of massless particles, even after their interaction is
effectively switched-off. We observe that by perfect black body spectrum of 
CMBR. However, this is not exactly true for neutrinos. As we mentioned 
above, neutrinos decoupled from cosmic plasma at ${ T = 2-3}$ MeV.
At that moment the primeval plasma consists of four weakly coupled parts:
electromagnetic ($e^+,e^-,\gamma$) and three separate neutrino parts.
At $T \sim $ MeV the temperature of photons became larger than the
temperature of neutrinos due to the heating of photons by $e^+e^-$-annihilation.
Residual annihilation of  hotter $e^+e^-$-pairs into neutrinos heats up
neutrinos and distort their spectrum~\cite{ad-fuk}. For more details and
references see review~\cite{nu-rev}.  
This leads to a larger neutrino number density, which can be described
as an increase on the effective equilibrium number of neutrino species:
\be
\Delta N_\nu = 0.03 + 0.01
\label{Delta-N}
\ee
Here the second term, 0.01, comes from the plasma corrections which
diminish ${ n_\gamma}$ with respect to the ideal gas
approximation~\cite{turner-plasma}. Because of these corrections the
number in eq.~(\ref{sigma-n-nu}) is not 336/cm$^3$, i.e. three times of 
eq. (\ref{nu-to-gamma}), but slighty larger.

This correction has a negligible impact on primordial $^4 He$ abundance, 
because there are two effects of the opposite sign.
An increase of the effective number of neutrino species, $\Delta N>0$,
leads to an increase of the mass fraction of the produced helium. On the other
hand, the account of the spectrum distortion of $\nu_e$ and their larger
number density has an opposite effect on $^4 He$ and the net effect is
quite small. 

However a larger energy density of the relic neutrinos has an impact 
on the form of the angular spectrum of CMBR which may be noticeable in the 
future CMBR missions, in particular, in Planck. The observation of this 
effect will be a measurement of physical processes which took place,
when the universe was only 1 second old.

\section{Neutrino and large scale structure formation \label{s'-nu-lss}}

Investigation of the large scale structure of the universe is an essential
ingredient of the modern cosmology. Comparison of the theory with observations
allows to study the properties of dark matter, measure cosmological parameters
and the spectrum of primordial density perturbations. In 
particular, an analysis of LSS permits to put more stringent than (\ref{m-nu-a})
bounds on neutrino mass. Bound (\ref{m-nu-hdm}) is already based on the
consideration of LSS but a more detailed study leads to stronger bounds.

\subsection{Basics of LSS  formation theory\label{ss-LSS}}

Theory of LSS formation is based on the following input:\\
1. {\it Spectrum of primordial density fluctuations}. The Fourier
transform of the power spectrum, i.e. of $(\delta \rho)^2$ is usually
parametrized in the simple form:
\be
\left(\frac{\delta \rho}{\rho}\right)_k \sim k^n, 
\label{delta-rho-k}
\ee
Observational data indicate that $n \approx 1$, i.e. the spectrum is
close to the so called flat, or Harrison-Zeldovich, spectrum. For such
spectrum metric fluctuations, which are dimensionless, do not contain
any dimensional parameter,
\be
\langle h^2_k\rangle \sim \int d^3k/k^{4-n}\sim \int dk/k
\label{h-k}
\ee
This type of spectrum is predicted by inflation with $n$ slightly different 
from unity. 

At large scales, from Gpc down to about 10 Mpc, the spectrum is measured 
by the angular fluctuations of CMBR. At the other end, from Mpc up to tens Mpc
the spectrum is measured by the observed LSS. A good test of the theory is that
these two kinds of independent measurements give coinciding results in the common
region of wave lengths around 10 Mpc.\\
2. {\it Type of the density perturbations.} It is assumed that the density
perturbations are the so called adiabatic ones, i.e. they can be visualized
as time-shifted in one space point with respect to another, $\rho = \rho (t(x))$,
as predicted by inflation. In other words, density perturbations are the same
in all forms of matter: baryons, dark matter, photons, etc. 
This type of perturbations is confirmed by CMBR at large scales, above 10 Mpc.
It is not excluded that at smaller scales the so called isocurvature perturbations
may be significant, but usually they are neglected. 

{\it Comment:} density perturbations at the early stage can be decomposed into
two independent modes: adiabatic and isocurvature. We have already defined adiabatic.
The isocurvature ones can be understood as perturbations in the chemical
content of the primordial matter and initially they have vanishing energy 
density perturbation, $\delta \rho =0$. Later, when the equation of state
becomes different due to different matter content the isocurvature perturbations
give rise to the normal density perturbations, $\delta \rho \neq 0$.\\
3. {\it Properties of dark matter}. According to observations the total energy
density of the universe is close to the critical one, 
\be
\rho \approx \rho_c = \frac{3H^2 m_{Pl}^2}{8\pi} \approx
10^{-29}\,\, {\rm g/cm}{ ^3}
\label{rho-c}
\ee
In terms of the dimensionless cosmological parameter $\Omega_j =\rho_j /\rho_c$
we have:
\be 
{ \Omega_{tot} = 1 \pm 0.05}, \,\,\,
{ \Omega_{B} \approx} 0.05,\,\,\,
{ \Omega_{DM} \approx} 0.25,\,\,\,
{ { \Omega_{DE} \approx}  0.7.}
\label{all-Omega}
\ee
where $B$, DM, and DE are respectively baryons, dark matter, and dark energy. 

Usually it is assumed that the cosmological dark matter consists of 
practically non-interacting cold dark matter (CDM) particles. Dark energy remains
mysterious but according to observations its equation of state is quite
close to the equation of state of vacuum energy, $p=-\rho$, and hence
dark energy is either vacuum energy (i.e. Lambda-term) or something
similar to it. So the standard model is called CDM$\Lambda$ model.
Still it is not excluded that dark matter may be self-interacting as e.g.
mirror matter and dark energy may have equation of state which varies with
time, $p = w(t) \rho$.\\
3. {\it Analytical calculations} at
linear regime, when ${\delta\rho/\rho\ll 1}$.
The calculations are done in the frameworks of classical 
physics: general relativity and hydrodynamics. The results are not distorted by
subsequent evolution at large scales, larger than 10 Mpc, 
accessible to CMBR.\\
4. {\it Numerical simulations} at nonlinear regime, when 
${\delta\rho/\rho\geq 1}$. The simulations are necessary at smaller 
scales, ${ \leq 10}$ Mpc. Simulations are usually done with dark matter
particles with masses of about $10^6 M_{\odot}$, where 
$M_\odot = 2\cdot 10^{33}$ g is the solar mass. Some essential physics 
in the simulations may be omitted especially at smaller scales
but the overall picture seems to be fine.

If any of the above is not true, the strong bound
on ${m_\nu}$ may be relaxed. On the other hand, all above is confirmed 
by the data. One cannot say that all the assumptions are strictly
proven, so some freedom remains, but the agreement of this simple
picture with the observations is quite impressive.

\subsection {\bf Neutrino role in LSS formation \label{ss-nu-role}}

Neutrinos stopped to interact with the primeval plasma at $T\gg m_\nu$
i.e. they were relativistic at decoupling. After that they freely propagate
in the universe with the red-shifted momentum.
Their free streaming length before they become nonrelativistic is huge:
\be
l_{FS} \approx 250\,{\rm { Mpc}}\,\,{({\rm { eV}}/m_\nu)} 
\label{l-FS}
\ee
The mass inside the free-streaming volume is
\be
M_{FS} = \frac {4\pi l_{FS}^3 \rho}{3} \approx 0.1\frac{m_{Pl}^3}{m_\nu^2}
\approx 10^{17} M_\odot ({\rm eV}/m_\nu)^2
\label{M_FS}
\ee
In neutrino dominated universe the structures below this scale should be 
erased. If neutrinos are sub-dominant they are still able to inhibit the structure 
formation at small scales. 

In the universe dominated by neutrinos very large structures with the size of the
order of $l_{FS}$ should be formed first and after that smaller structures could be
formed by fragmentation of large scale structures to smaller ones. In other words
smaller size structures are younger. The same effect but less pronounced would exist
in mixed hot-cold dark matter scenario. Thus the larger is the neutrino mass 
(but still small) the longer would be the delay in the formation of small structures.
In particular, the number of Ly-${\alpha}$ clouds 
at high red-shifts, ${z\geq 1}$, is very sensitive to light massive neutrinos.
The data analysis permits to restrict the neutrino mass by:
\be {
\sum m_\nu < 0.7\,\,{\rm eV} },
\label{m-nu-LSS}
\ee
i.e. the individual neutrino mass is bounded by $m_\nu < 0.23 $ eV.

On the other hand, formation of Lyman-${\alpha}$ clouds is sensitive to 
gasodynamics and possible shock waves which are not yet included into numerical
simulation at the non-linear stage of perturbation evolution and
are {poorly understood.} Another unresolved problem is that of biasing, i.e.
of the shift between the distributions of the visible and dark matter. The 
biasing is observationally determined at galactic scales or larger and is
extrapolated to smaller scales but it is unclear how much it changes with
the size of the object.

An ad hoc modification of perturbation spectrum {at intermediate scales}
{could mimic presence of massive neutrinos and} 
{a less restrictive mass bound from LSS would be obtained.}

On the other hand, a good agreement between perturbations at the 
scales accessible to the measurements of 
angular spectrum of CMBR, especially by WMAP, and smaller 
scales, measured in the analysis of LSS by 2dFGRS and SDSS, 
in the common region, accessible to both CMBR and LSS,
demonstrates that  the picture is self-consistent.

For a review of different methods of determination/restriction of the neutrino
mass from the astronomical data see ref.~\cite{m-nu-rev}. 

\section{ Neutrino mass and CMBR \label{nu-m-CMBR}}

The spectrum of the angular fluctuations of CMBR provides us with accurate 
information on cosmological parameters and, in particular, on neutrino mass
and on the number of neutrino species. Theoretically calculated spectrum of
CMBR angular fluctuations for the accepted values of the cosmological
parameters is presented in fig. 1. The shape of the spectrum depends upon
neutrino mass and its analysis allows to put a stringent bound on 
the sum of the masses of all three neutrino species. In fig. 1,
taken from ref.~\cite{ichik} five curves corresponding to different neutrino
masses $m_\nu = 0.5,\,\, 0.7,\,\, 1.0,\,\, 2.0$ eV are presented and compared
with the observational data from WMAP3.

{Massive neutrinos lead to the following two effects in the angular fluctuations of 
CMBR:\\ 
{1. Shift of peaks to the left with rising mass. The larger 
${ m_\nu}$, the earlier is the non-relativistic stage. Thus the
distance to the last scattering surface is shorter and peaks move
to smaller ${l}$}.
However, this effect can be compensated by a shift in $ H$.\\
{2. Decrease of the (1st) peak hight.}
{Neutrinos with ${m_\nu > 0.6}$ eV become 
nonrelativistic before recombination,} 
so the matter radiation equality takes place earlier
and the enhancement of the first peak by the integrated Sachs-Wolfe
(ISW) effect becomes weaker.
\begin{figure}
\includegraphics{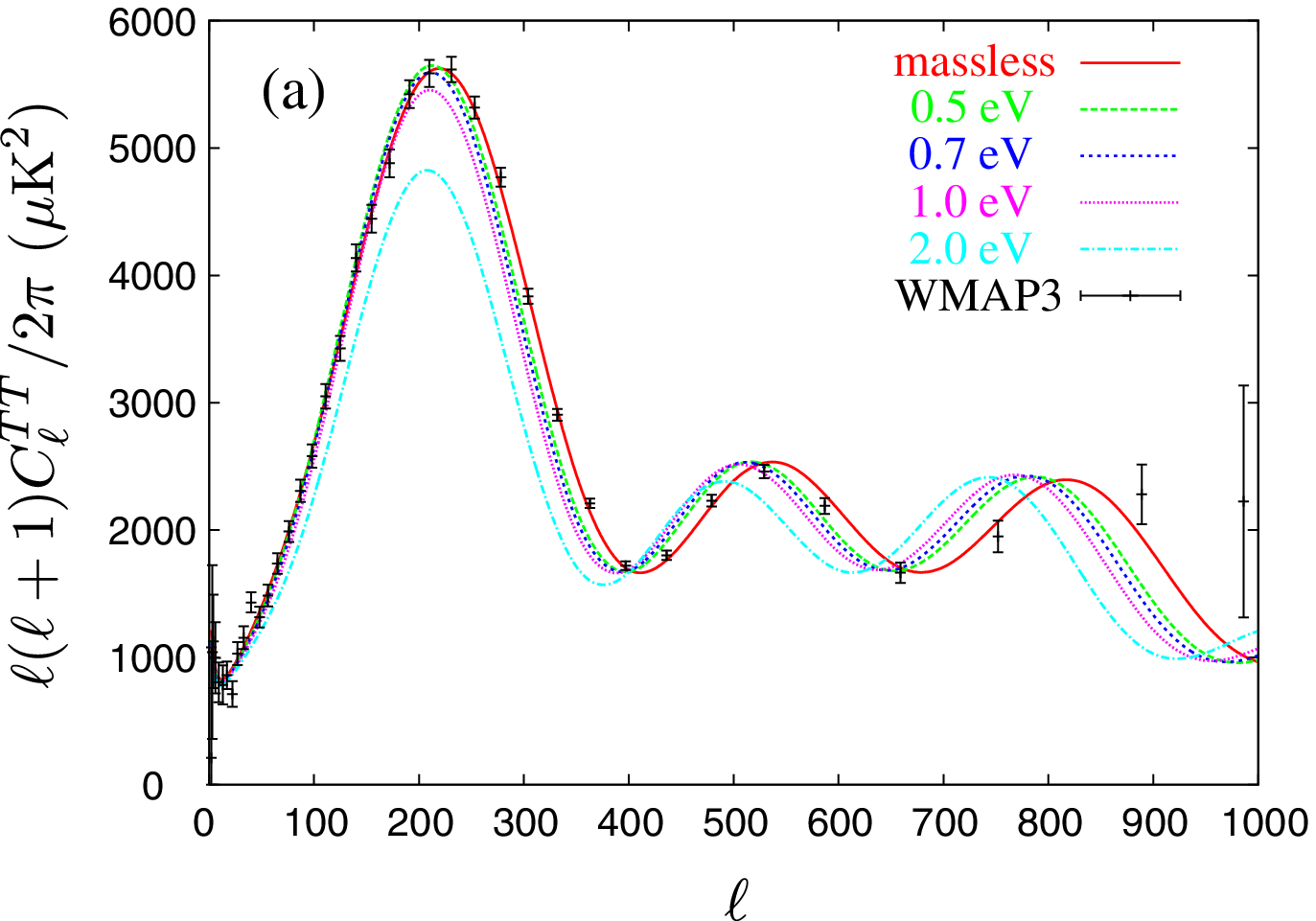}
\end{figure}
$$ $$
$$ $$ 
$$ $$
$$ $$
$$ $$
$$ $$
$$ $$
$$ $$
$$ $$
$$ $$
$$ $$
{Figure 1. Effects of massive neutrinos on the spectrum of angular fluctuations
of CMBR. The curves from top to bottom in the first maximum correspond to
$m_\nu =0,\,\, 0.5,\,\, 0.7,\,\, 1.0$, and $2.0$ eV}.
\\[2mm]

The degeneracy of the effects induced by the neutrino mass and
the magnitude of the Hubble parameter is illustrated in fig. 2~\cite{ichik}.
On the same figure the measured values of $H$ by three different groups are
presented. 

The analysis of WMAP data allows to restrict the neutrino mass from above
by~\cite{ichik}:
\be
m_\nu < 0.63\,\,eV
\label{m-nu-WMAP}
\ee
at 95\% confindence level. The derivation does not use data on LSS which
are subject to ambiguities due to poorly understood biasing and nonlinearity 
in structure evolution. 
$$ $$
$$ $$
$$ $$
$$ $$
$$ $$

\begin{figure}[t]
\includegraphics{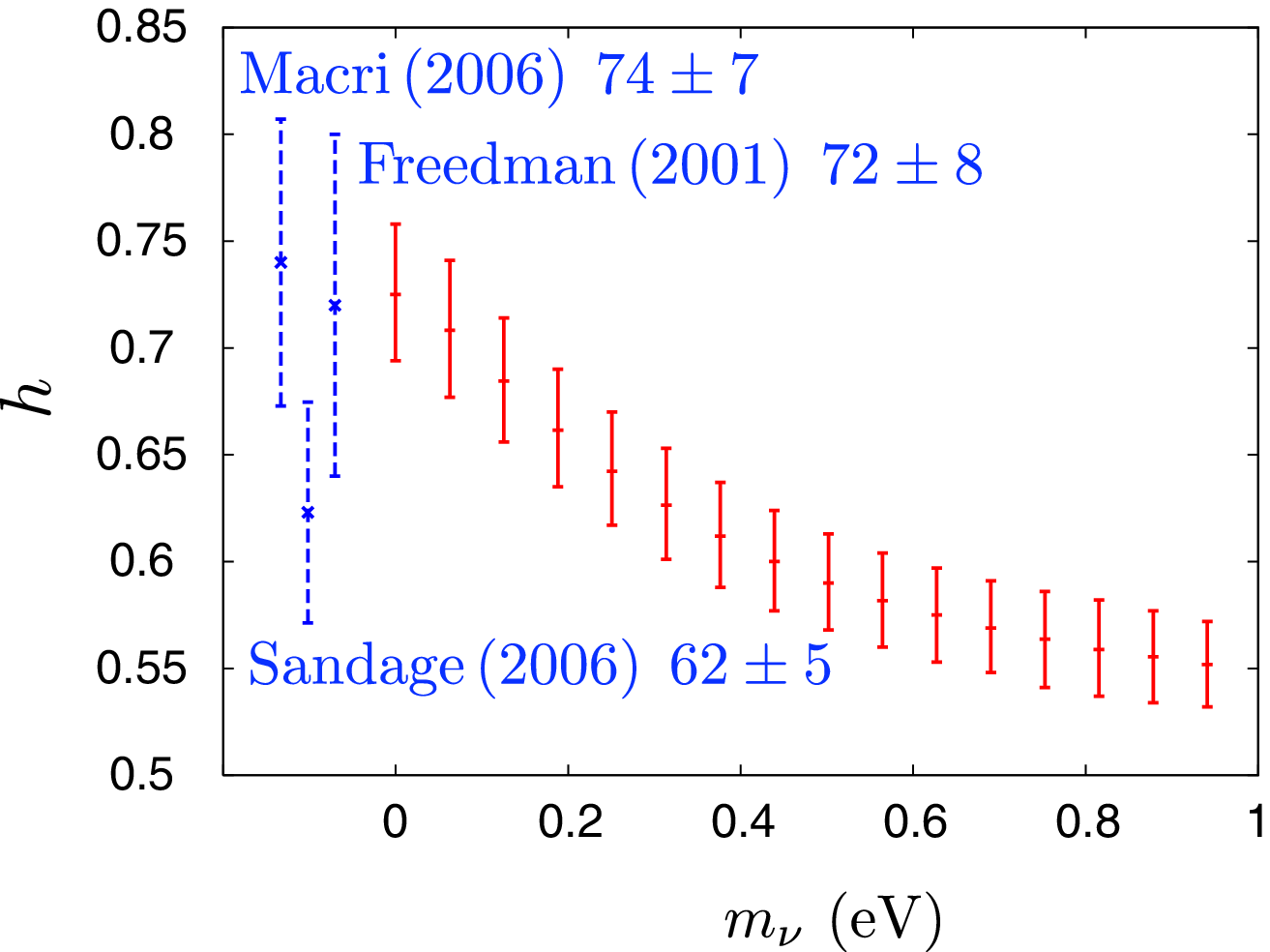}
\end{figure}
$$ $$
$$ $$
$$ $$
$$ $$
$$ $$
$$ $$
$$ $$
$$ $$
Figure 2. The constraints on h for several fixed values of neutrino 
mass. The first three bars at the left hand side are the constraints from 
distance ladder measurements.

\section{Massive conclusion \label{s-mass-cncl}}

Here we summarize \\
1. All bounds quoted above are based on the calculations of the present
day number density of neutrinos initiated by GZ. \\
2. From CMBR only it follows: ${m_\nu < 0.63}$ eV.\\
3. More restrictive bounds obtained in several papers under
different assumptions vary in the interval:
${ \Sigma\mnu <1-0.42}$ eV. These results are based on the
analysis of LSS at relatively small scales plus CMBR to shift degeneracy 
of parameters.\\
4. With almost equal neutrino masses
one can conclude: ${{ \mnu < 0.63-0.14}}$ eV,
lower value is probably too strong.\\
4. The planned detector KATRIN, expects to reach the accuracy 
${ m_\nu < 0.35}$ eV at $ {5\sigma}$. Do we need this expensive experiment?\\
5. Possible ways to violate the upper limits on $m_\nu$: \\
a) From particle physics: new particles or stronger than standard
${\nu}$-interactions;\\
b) From cosmology: unusual spectrum of (isocurvature) perturbations at small 
scales or low (re)heating temperature. 

So to conclude, we need KATRIN. The issue of neutrino mass is too important
to rely only on astronomical measurements. However, it is quite probable that
the absolute value of the neutrino masses is close to their mass differences.
In this unfortunate case KATRIN will not help but there may be a chance 
to weight neutrinos by the future Planck mission.

\section{Detection of cosmic neutrino background (C$\nu$B)
\label{s-C-nu-B}}

Though the number density of cosmic neutrinos is very high, it is not yet
directly registered. There are some indirect ways to ``observe'' C$\nu$B,
for example by its impact on LSS and CMBR in the case of
sufficiently large $m_\nu$, as is discussed above. Among other possibilities
there are:\\
1. {\it  Measurement of the number of neutrino families  through BBN.} 
Larger number of {equilibrium} neutrino species shifts ${n/p}$-freezing to 
higher ${T}$ and shortens the time of nuclear formation. One neutrino 
family makes $ {^4He}$ 5\% larger. With ${n_B/n_\gamma}$ known from CMBR:
\be 
{N_\nu = 3\pm 0.3}
\label{N-nu-BBN}
\ee
(for more detail see below sec.~\ref{s-BBN}).\\
2. {\it The angular spectrum of CMBR.} A change of the number and energy density
of the relic neutrinos would shift the moment of matter-radiation equality, i.e.
the moment when the universe stopped to be dominated by relativistic matter
and the cosmological expansion regime changed into the matter dominated one.
This would shift the positions of peaks of CMBR spectrum. The effect is rather 
weak and the precision is lower than that of BBN.
For massless or very light neutrinos the number of effective neutrino families
is~\cite{n-nu-cmbr}
\be
{N_\nu = 3\pm 1}
\label{N-nu-CMBR}
\ee
It is clearly a long way to the necessary level of accuracy to observe the
effect discussed in sec.~\ref{ss-noneq-nu}, eq. (\ref{Delta-N}).\\
3. {\it Direct detection of C${\nu}$B:}\\
a)  The evident process of registration of the relic
neutrinos by ${ \nu N}$-scattering seems to be out of question because
the corresponding cross-section is extremely small: 
\be 
\sigma_{\nu N} \sim G^2_F E_\nu^2 \sim 10^{-55}\,cm^2\,
\left(\frac{E_\nu}{eV}\right)^2
\label{sigma-nu-N}
\ee
Since the average energy of massless cosmic ${\nu}$ is
${ \langle E \rangle \approx 3.15 T \approx 5.3 \cdot 10^{-4}}$ eV we obtain
$\sigma_{\nu N} \approx 10^{-62}$ cm$^2$. Even for massive neutrinos the
situation is not much better.\\
b) $ Z$-burst effect~\cite{Z-burst} opens a more realistic chance to register
C$\nu$B. To this end a flux of ultrahigh energy cosmic ray neutrinos is
necessary. The $Z$-resonance scattering of the high energy neutrinos on the 
background ones has much larger cross-section:
\be
\sigma_Z \sim \alpha/m_Z^2 \approx 10^{-33}-10^{-34} \,\,{\rm cm^2}
\label{sigma-Z}
\ee
However, to excite the $Z$-resonance on massless $ \nu$ with 
${E \sim 10^{-3}}$ eV  very high energy of cosmic ray
neutrinos is necessary:
\be {
E_{\nu;hi} = 4\cdot 10^{21}\,\, eV \left(eV/E_{\nu;low}\right)
}\label{E-nu-hi}
\ee
Even if neutrinos with $E\sim 10^{24}-10^{25}$ eV exist in the cosmic rays,
their flux most probably is negligibly small and the chance to register the Z-burst is
practically zero, but it may be feasible for massive, not too light $\nu$.
\\
c) The cross-section of elastic scattering of low momentum Dirac neutrinos 
can be strongly enhanced by coherent effects. Neutrinos efectively interact with
all nuclei inside the volume $\lambda^3$, where $\lambda = 1/m_\nu v $ is the inverse
neutrino momentum. Correspondingly the cross-section would be larger than that
for a single nucleus by the factor $N^2$, where $N$ is the number of nuclei  
inside $\lambda^3$. According to the estimates of ref.~\cite{schw} the acceleration
of a small piece of matter with the size $l\sim \lambda$ could reach the value
$10^{-22}$ cm/s$^2$. This result was obtained for $m_\nu = 30$ eV and the galactic
density of neutrinos $n_{(gal)} = 10^7$/cm$^3$. In other words, it was assumed
that neutrinos make all dark matter in the galaxy. Now we know that neutrinos 
are much lighter and can make at most 15\% of all dark matter. Correspondingly the
effect should be 6 times weaker or maybe even more than that if light neutrinos
are not accumulated in the galaxy in the same fraction as cold dark matter.
\\
d) An interesting possibility is open by the inverse beta decay, induced by 
the relic neutrinos:
\be
\nu + A \rar A' + e,
\label{inv-beta}
\ee
This process should take place if nucleus
$A$ $ \beta$-decays into nucleus $ A'$. The reaction has zero
threshold but if ${m_\nu = 0}$ the spectrum of the produced electrons 
in the inverse beta decay very 
weakly deviates from the spectrum of electrons in the beta-decay.
As was suggested in ref.~\cite{sw-beta}, 
the spectrum distortion may be observed if the chemical potential of C$ \nu$B
is large, ${ \mu_\nu \geq T_\nu}$. However, BBN plus large mixing angle solution
to the solar neutrino anomaly strongly restricts chemical potential of all
neutrino flavors~\cite{mu-limit} (see sec.~\ref{s-BBN}): 
\be
\mu_\nu < 0.07 T_\nu
\label{mu-nu}
\ee
The $ \beta$-capture of massive neutrinos~\cite{cmm} looks more promising
for registration of C$\nu$B, if $m_\nu$ is not too small.
The allowed kinematical region for the electron energy in
$\beta$-decay is ${ m_e<E_e<W_0}$, while in the 
${ \nu}$-capture it is $ {E_e > W_0 +2m_\nu}$.  According to the calculations
made in ref.~\cite{cmm}, the ratio of the reaction rates for the detector 
resolution $ \Delta$ is equal to:
{\be {
\frac{\lambda_\beta(\Delta)}{\lambda_\nu} = \frac{2}{9 \zeta(3)}
\left( \frac{\Delta}{T_\nu} \right)^3 \left(1+ \frac{2
m_\nu} {\Delta} \right)^{3/2} >> 1
}\label{lambda-beta}
\ee}
where ${ T_\nu \approx 1.7\cdot 10^{-4}}$ eV.

The above estimate (\ref{lambda-beta}) is valid for the conventional C$\nu$B 
with the distribution function given by eq.~(\ref{f-nu})
but if by some reason there are additional contributions into the
number density of the cosmic neutrinos the result may be quite different. 
In this connection is tempting to ask if
the events observed in experiment~\cite{lobashev} out of the kinematically allowed
region of the tritium beta-decay be explained by some anomalous
contribution into $f_\nu$? 

According to the calculations of ref.~\cite{cmm} the neutrino capture rates
in events/year, for 100 g of $^3$H, for the Fermi-Dirac distribution (\ref{f-nu})
over neutrino momentum but not energy, 
with ${ T_\nu =1.7\cdot 10^{-4}}$ eV (FD), for  
a Navarro-Frenk-White profile (NFW), and for the present day mass 
distribution of the Milky Way (MW) are respectively: 
7.5 (FD), 23 (NFW), and 33 (MW) if $m_\nu = 0.3 $ eV and
7.5 (FD), 10 (NFW), and 12 (MW) if $m_\nu = 0.15 $ eV.

\section{Neutrino and big bang nucleosynthesis (BBN) \label{s-BBN}}

Big bang nucleosynthesis took place when the universe was quite young,
${ t = 1-300}$ sec, and not too hot, ${ T= 1 - 0.07}$ MeV. At this
period primordial light elements,  $^2H$, $^3He$,  $^4He$, and  $^7Li$,  
were created.
Physics in this energy range is pretty well known and the results of 
calculation are in a good agreement with the observations. Deviations
from the usual physics would destroy the agreement between theory
and the observational data, so BBN serves as an efficient ``cleaner'' of
distortions of the standard physics.

The building blocks for formation of light elements were prepared by 
the reaction of neutron-proton transformations:
\be{{
n + e^+ \lrar p + \bar\nu_e}}
\nonumber\\
{{ n + \nu_e \lrar p + e^-}}
\label{n-p-tran}
\ee
These reactions were frozen at ${ T_f\approx 0.65}$ MeV,
and this determines the starting value of ${ n/p}$-ratio,
$n/p \approx \exp (-\Delta m_{np} /T_f) \approx 0.135$.
When ${T}$ drops down to 60-70 keV (it corresponds roughly to 200 sec) 
all neutrons, which survived the decay, quickly form ${ ^4 He}$ (about
25\% by mass) and a little ${ ^2H}$ (${ 3\times 10^{-5}}$ by number),
${ ^3 He}$ (similar to ${ ^2H}$) and 
${ ^7 Li}$ (${ 10^{-9}-10^{-10}}$). 

It is already clear from reactions (\ref{n-p-tran}) that neutrinos play
an important role in BBN. There are several neutrino effects which influence
the primordial abundances of light elements:\\
1. {\it Impact of neutrinos on the cosmological cooling rate.} The energy
density of the primeval plasma is: 
\be
\rho = \frac{3m_{Pl}^2}{32\pi t^2} = \frac{\pi^2}{30}\,g_* T^4 
\label{rho-of-T}
\ee
where
\be{{
g_* = 10.75 + 1.75 \Delta N_\nu  }
}\label{g-star}
\ee
is the number of the degrees of freedom of particles which were present in
the plasma. The first term, 10.75, comes from photons, $e^\pm$-pairs, and
three families of left-handed neutrinos. All the rest (which is absent in the
standard model) is parametrized as $\Delta N_\nu$. This 
${ \Delta N_\nu}$ includes {any form of energy} present at BBN. $\Delta N=1$
corresponds to energy density of equilibrium neutrinos and antineutrinos with
vanishing chemical potential and one polarization state.
In particular, if neutrinos are degenerate, i.e. $n_\nu \neq n_{\bar \nu}$
and thus their chemical potential is non-zero their energy density would be 
larger than the usual equilibrium one by:
\be{{
\Delta N_\nu = \frac{15}{7}\left[\left(\frac{\xi}{\pi}\right)^4+
2\left(\frac{\xi}{\pi}\right)^2\right]
}}\label{N-BBN}
\ee
where ${{ \xi =\mu/T}}$.

Positive ${\Delta N_\nu}$ leads:\\
1) to an earlier ${ n/p}$-freezing and 
higher frozen ${ n/p}$-ratio;\\
2) to a faster cooling down to the BBN temperature, 
${ T_{BBN}} \approx 65 $ keV, and to 
higher ${ n/p}$ as well, because less neutrons would have chance to decay.
The net result is an increase of ${ ^4 He}$ by about 5\%.

Negative $\Delta N$
leads to sign-opposite results. From the observed abundances of $^4 He$
and known from CMBR ratio of baryons to photons, 
$n_B/n_\gamma =6\cdot 10^{-10}$ one can conclude that
$
\Delta N_\nu < 0.3,
$
though, strangely, $\Delta N_\nu <0$ or {${ N_\nu < 3}$} seems most favored.
A recent analysis of determination of $n_B/n_\gamma$  and $\Delta N$ from
BBN and CMBR can be found in ref.~\cite{si-st}. \\
2. {\it Impact of degeneracy of electronic neutrinos.} If chemical potential
of $\nu_e$ is non-zero, its effect on BBN is much stronger than the effect of
comparable chemical potentials of $\nu_\mu$ and $\nu_\tau$. A non-zero $\xi_e$
shifts equilibrium value of ${ n/p}$-ratio as:
\be
\left( n/p \right)_{eq} = \exp \left( -\frac{\Delta m_{np}}{T} -\xi_{e}
\right)
\label{n-p-eq}
\ee
Hence the bounds on the chemical potentials of $\nu_e$ and $\nu_{\mu,\tau}$
are very much different:
\be{{ 
|\xi_{\mu,\tau}| < 2.5}} \label{xi-nu-mu-tau}\\
{ |\xi_{e}| < 0.1.
}\label{xi-bounds}
\ee
They are valid 
if  a {compensation} between the effects induced by ${{ \xi_{\mu,\tau}}}$ 
and ${ \xi_e}$ is allowed. In the absence of such compensation the bounds are 
somewhat stronger. We will see below, however, that the bounds on chemical 
potentials of all flavors are much stronger because of efficient neutrino
oscillations in the early universe.\\
3. {\it Variation of the energy density of electronic neutrinos.}
If the asymmetry between neutrinos and antineutrinos is negligible 
but the energy density of $\nu_e$ differs from the equilibrium one, the
effect of that would have the opposite sign with respect 
to the effect induced by
non-zero $\Delta N_\nu$. If {${ n_{\nu_e} > n_{eq}}$}, the
${{n-p}}$ freezing temperature would go down and so would
the ${ n/p}$-ratio. In the opposite case, if ${ n_{\nu_e} < n_{eq}}$} 
the ratio ${ n/p}$ would rise.\\
4. {\it The effect on BBN of the oscillations between active neutrinos in equilibrium} 
is negligible because in this situation the oscillations do not change 
anything in the plasma. However, if neutrino asymmetries are nonvanishing, the
oscillations, which do not respect conservation of the flavor lepton
numbers, lead to transitions between electronic, muonic, and tauonic 
asymmetries. We know that the large mixing angle solution to Solar 
neutrino deficit is realized. Hence initially large $\xi_\mu$ or $\xi_\tau$
would be effectively transformed into $\xi_e$, prior to BBN
and all chemical potential would be equalized. Since $\xi_e$ is strongly 
restricted, similar bounds can be imposed on all $\nu$-flavors~\cite{mu-limit}:
\be
|\xi_a | < 0.07
\label{xi-a}
\ee
Evolution of lepton charge asymmetry for LMA solution 
to solar anomaly is shown in fig.~3.

Such small neutrino degeneracy cannot have a noticeable 
cosmological impact, in particular, on LSS and CMBR.
However, bound (\ref{xi-a}) can be relaxed~\cite{ad-ft}, if neutrinos have 
a new interaction with light scalar bosons, Majorons. The potential of neutrinos 
induced by the Majoron exchange inhibits early oscillations in the plasma.
\begin{figure}[ht]
\includegraphics{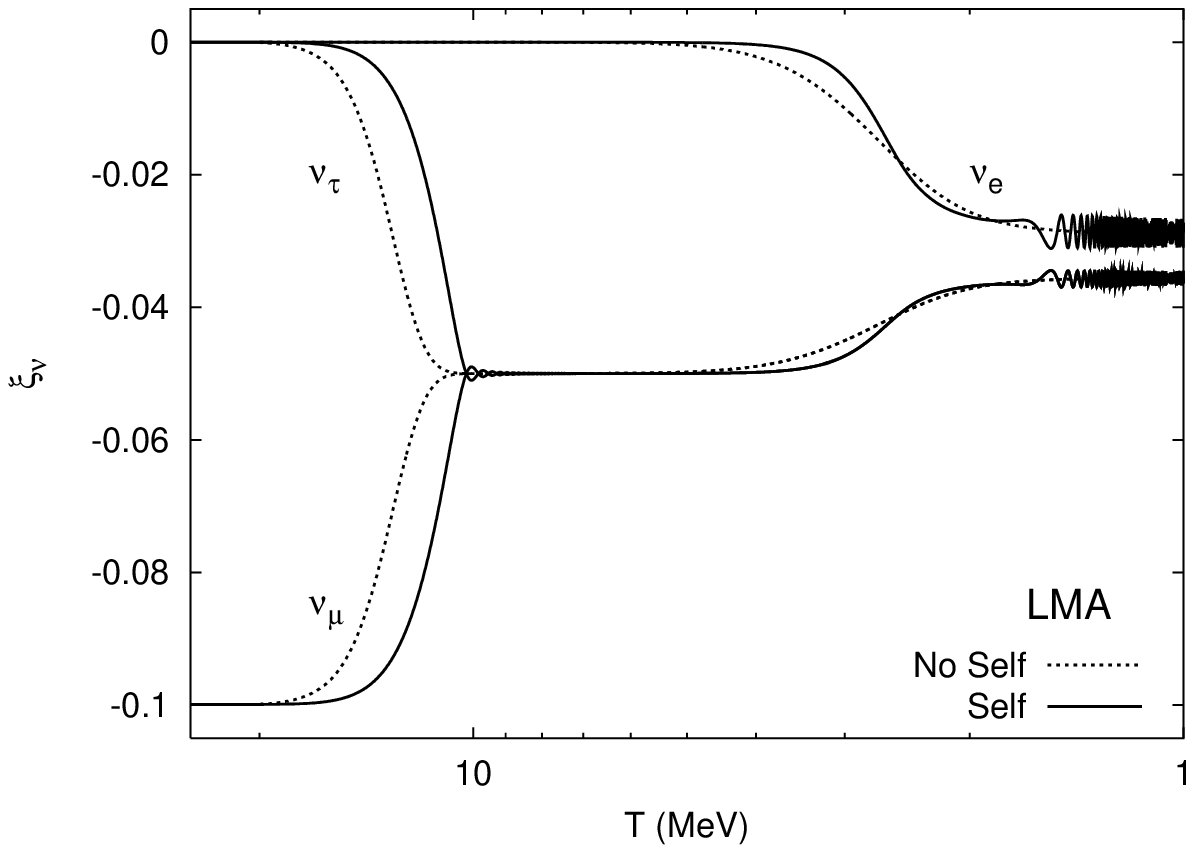}
\end{figure}
$$ $$
$$ $$
$$ $$ 
$$ $$
$$ $$
$$ $$
$$ $$
$$ $$
$$ $$
$$ $$
$$ $$
$$ $$
Figure 3. Transformation of initial chemical potential of $\nu_\mu$ 
into those of $\nu_e$ and $\nu_\tau$ as a function of the plasma temperature,
with and without neutrino self-interactions. \\[4mm]

\section{Cosmological impact of neutrino statistics violation
\label{s-stst-viol}}

According to the famous Pauli theorem, particles with
integer spins obey Bose-Einstein statistics, while those with
half integer spin obey Fermi-Dirac statistics. There is no consistent
formulation of a theory where this theorem is violated. Nevertheless
phenomenological manifestations of spin-statistics violation are 
discussed and experimental bounds are presented. Such bounds are very 
strong for electrons and nucleons, while for neutrinos a large
spin-statistics violation is allowed. Such a possibility is discussed
in ref.~\cite{ad-as}. In this paper one can also find a list of references
to theoretical and experimental works on the violation of the Pauli 
theorem. 

The best experimental bound on breaking of the exclusion principle for 
neutrinos follows from the double beta decay but still even this best 
bound allows for about 50\% violation~\cite{bddss}. In other words, in principle
neutrino may be half and half of fermionic and bosonic parts.
In fact, the existing data allow neutrinos to be more than half bosonic,
up to 60-70\%.

If neutrino statistics is not purely the Dirac-Fermi one, there may be
very interesting cosmological consequences. Evidently the theoretical
predictions for abundances of light elements created at BBN would 
change~\cite{nu-b-BBN}. 
If neutrinos were purely bosonic, their energy density would be 8/7 of 
normal fermionic $ \nu$, giving $ {\Delta N_\nu = 3/7}$. On the other 
hand, a larger number density of $\nu_e$ would lead to smaller temperature 
of ${n/p}$-freezing and thus is equivalent to negative $\Delta N_\nu$.
The net result is that the effective number of neutrinos at BBN should
be smaller than three: {${ {N_\nu^{(eff)} =  2.43}}$}. This is exactly the 
number obtained from the data analysis in ref.~\cite{si-st}.

In ref.~\cite{nu-b-BBN} it was argued that the equilibrium neutrino
distribution for mixed statistics has the form:
\be
f^{(eq)}_\nu = \left[ \exp (E/T) + \kappa \right]^{-1}.
\label{f-mixed}
\ee
where the parameter $\kappa$ changes from (-1) to (+1) and interpolates between 
Bose and Fermi statistics.

Another interesting effect of (partly) bosonic neutrinos is that 
they might form cosmological Bose condensate. In the (excluded) case of 
purely bosonic neutrinos their equilibrium distribution would be
\be 
f_{\nu_b} = \frac{1}{\exp [(E-\mu_\nu)/T -1] } + C \delta^{(3)} (k),
\label{f-nu-b}
\ee
where the chemical potential reaches the maximum allowed value
for bosons: $\mu_\nu^{(max)} = m_\nu$. If the equilibrium distribution of
neutrinos is mixed one~(\ref{f-mixed}), with $(-1)<\kappa<0$
the maximum value of the chemical potential, at which Bose condensation
of neutrinos could take place, would be $\mu_\nu^{(max)} = m_\nu -T\ln|\kappa|$.

If neutrinos might Bose condense then they can make not only hot dark
matter but also (all) cold dark matter and we do not need for that
any new particles or other objects. One can avoid the lower bound on
neutrino mass~\cite{tr-gunn} based either on Fermi statistics or (for
bosons) on the Liuville theorem~\cite{mad} because of the condensate
formation. So the cosmological dark matter can 
consist of old known particles (massive, though very light neutrinos) 
but to this end we need very new physics. 

If the neutrino condensate makes all cosmological dark matter, 
its average number density should be about $10^4(m_\nu/0.1\,{\rm eV})$/cm$^3$,
while in the Galaxy it can be 5-6 orders of magnitude larger.
Probably it is not enough to explain the observation of electrons 
out of the kinematically allowed region in the tritium beta decay,
mentioned at the end of sec.~\ref{s-C-nu-B}.

\section{Neutrino oscillations in the early universe \label{s-nu-osc}}

Here we will consider possible effects on BBN if the usual active neutrinos
are mixed with new sterile ones. Neutrino propagation in plasma is governed
by the effective potential (refraction index)~\cite{eff-pot-nu}:
\be
V_{eff}^a =
\pm C_1 \eta G_FT^3 + C_2^a \frac{G^2_F T^4 E}{\alpha} ~,
\label{V-eff}
\ee
where the constants $C_j$ are of the order unity and ${\eta}$ is the plasma 
charge asymmetry:
\be
\eta^{(e)} &=&{{{{
2\eta_{\nu_e} +\eta_{\nu_\mu} + \eta_{\nu_\tau} +
\eta_{e}-\eta_{n}/2 \,\,\,\,
 ({\rm for}\,\,\,  \nu_e)}}}
}\label{eta-e} \\
{{\eta^{(\mu)}}} &=& {{
2\eta_{\nu_\mu} +\eta_{\nu_e} + \eta_{\nu_\tau} - \eta_{n}/2\,\,\,
({\rm for}\,\,\,\nu_\mu)}}
\label{eta-mu}
\ee
The first term comes from the averaging of the neutrino current
${\langle J_0 \rangle}$ over plasma. It has different signs for neutrinos
and antineutrinos. It is proportional to the charge asymmetry in plasma, 
i.e. to the difference of the number densities of particles and antiparticles
and in cosmological situation it is usually small but in stars or in the Earth
it gives the dominant contribution.
The second term originates from the non-locality of weak interactions 
related to the exchange of the intermediate bosons. Because of that neutrino
can be absorbed in one space-time point and re-emitted in different one.
This term is usually negligible in astrophysics but is dominant in cosmology
especially at high temperatures, ${ T\sim 10}$ MeV.

Mixing between active and sterile neutrinos would lead to the following
phenomena which in turn could produce noticeable effects on BBN:\\
1. An increase of the number of particle species due to production of sterile
neutrinos. It leads to ${{ N_\nu >3}}$.\\
2. Since the probability of the oscillations depends upon neutrino
energy, the neutrino spectrum could be different from the usual equilibrium 
one. As we mentioned above, BBN  is very sensitive to the spectrum
distortion of $\nu_e$. \\
3. Generation of large lepton asymmetry in the active neutrino
and especially in ${{\nu_e}}$  sector by the resonance oscillations
between active and sterile neutrinos~\cite{foot}.\\
For more details see review~\cite{nu-rev}

In the simplified case when the mixings of active neutrinos are neglected,
and the mass difference between active and sterile neutrinos is positive, 
so there is no resonance transition, the
bounds on the mixing angle and the mass difference between one active
and one sterile neutrino are:
\be
(\dm_{\nu_e\nu_s}/{\rm eV}^2) \sin^4 2\theta^{\nu_e\nu_s} =
3.16\cdot 10^{-5} \ln^2 (1- \Delta N_\nu)\\
\label{nu-e-mix}
(\dm_{\nu_\mu \nu_s}/{\rm eV}^2) 
\sin^4 2\theta^{\nu_\mu\nu_s} =
1.74\cdot 10^{-5} \ln^2 (1-\Delta N_\nu)
\label{nu-mu-mix}
\ee
 
The impact of oscillations between active and sterile neutrinos for
realistic mixing between active neutrinos for both resonance and
non-resonance cases was considered in ref.~\cite{ad-fv}.
Unfortunately the results cannot be presented in a simple analytical form, 
as above, but only as figures.

\section{BBN and right-handed neutrinos, $W_R$,  $Z_R$,
neutrino magnetic moments, etc \label{s-nu-R}}

We mentioned above that if neutrino mass is non-zero, there must be
both helicity states, right-handed and left-handed. Now let us estimate 
the rate of production of the ``wrong'' right-handed neutrinos in
the early universe. The most favorable period for the production
of $\nu_R$ in the standard theory, i.e. without right handed currents,
only by neutrino mass, took place at the temperatures of 
the order of $W$ and/or $Z$ mass, when these intermediate bosons
were present in the plasma. The production rate of $\nu_R$ by the decays
of $W$ and $Z$ is equal to:
\be
\Gamma_R =\frac{\dot n_{\nu_R}}{ n_{\nu}} \approx
10\,\left(\frac{m_\nu}{T}\right)^2 \,
\frac{\Gamma_W^\nu n_W + \Gamma_Z^\nu n_Z}{T^3},
\label{Gamma-R}
\ee
where $\Gamma^\nu_{W,Z}$ is the decay width of $W$ or $Z$ into neutrino
channel and $n_{W,Z}$ are the number densities of these bosons.
Comparing $\Gamma_R$ with the expansion rate, $H\sim T^2/m_{Pl}$,
one sees that $\nu_R$ were never produced abundantly if ${m_\nu}$ 
respected the GZ-bound. 

If in addition to the usual weak intermediate bosons, $W_L$ and $Z_L$ (we
omitted above the sub-index $L$) there exist right-handed bosons which
interact with right-handed fermions, the probability of interactions
of e.g. right-handed neutrinos would be inversely proportional to
$m_{W_R}^{-4}$. The energy density of $\nu_R$ at BBN should be smaller than,
say, 0.3 of the energy density of the usual left-handed neutrinos.
It demands that $\nu_R$ should decouple before the QCD phase transition.
If so, the right-handed neutrinos would be diluted to the necessary amount
by the entropy released in the massive particles annihilation. This condition
leads to the lower bound of $W_R$ mass:
\be
\frac{m_{W_R}}{m_{W_L}} > 2.5 \,\,{\rm TeV}\,\,\left(\frac{T_{QCD}}
{200 \,{\rm MeV}}\right)^{4/3}
\label{m-W-R}
\ee  
This is an order of magnitude better than the direct experimental limit.

There is a different model for generation of right-handed neutrinos
by the mixing between ${W_R}$ and ${W_L}$:
\be
W_1 = \cos\theta W_L + \sin \theta W_R 
\label{W-1}
\ee  
In this case the production of ${\nu_R}$ at ${T \leq T_{QCD}}$ 
is given by
\be
r=\Gamma_R/H = 
\sin^2\theta \left( \frac{T_{QCD}}{2\,{\rm MeV}}\right)^3  
\label{r}
\ee
Demanding ${r<0.3}$ results in 
$\sin^2 \theta < 10^{-6}$. {It is interesting that the effect does not vanish 
for ${{ m_{W_2} \rar \infty}}$.} 

The increased total number of neutrino species, which would lead to a larger
value of $(n/p)$-ratio, may be compensated by an increase of the number
of the right-handed electronic neutrinos which makes
the ${{ n-p}}$ transformation more efficient and thus diminish their frozen
ratio. However the interaction of $\nu_R$ with nucleons is proportional to
${\sim \sin^2\theta}$ and the effect is small, if $\theta < 10^{-3}$, as is 
written above.
On the other hand, a large ${\theta\sim 1}$ may be allowed because in
this case the compensation would be possible.

If neutrinos have a non-zero magnetic moment, $\mu_\nu$, it also leads
to production of right-handed helicity states. For example,
${\nu_R}$ could be produced in the electromagnetic reaction:
\be{{
e^\pm + \nu_L \rar e^\pm + \nu_R}
}\label{e-nu}
\ee
Demanding that right-handed neutrinos contributes not more than
 ${ \Delta N_\nu = 0.5}$, we obtain:
\be{{
\mu_\nu < 3\times 10^{-10} \mu_B},
}\label{mu-B-nu}
\ee
where $\mu_B$ is the Bohr magneton.

If there exists primordial magnetic field, coherent on sufficiently
large distances, right-handed neutrinos can be produced by spin flip in
this field. In this case the magnetic moment is bounded by:
\be
\mu_\nu < 10^{-6} \mu_B \left(B_{primord}/{\rm { Gauss}}\right)^{-1}
\label{mu-nu-lim}
\ee
The magnitude of the $B_{primord}$ is unknown but if we assume that the 
intergalactic magnetic fields have the strength 
${{ B_{int-gal} \sim 10^{-6}}}$ Gauss 
and that they were generated in the early universe and evolved adiabatically
to the present time, their strength at $T=1 $ MeV should be $10^{13}$ G and
we obtain a very  strong bound:
\be{{
\mu_\nu <  10^{-19} \mu_B
}} \label{mu-nu-lim2}
\ee

\section{Conclusion \label{s-concl}}

Thus we see that cosmology and astronomy happen to be  
the most efficient neutrino ``detector'' at the present time.
The upper bound on neutrino mass at the level of a fraction of eV
is noticeably stronger than that obtained by direct experiments.
On the other hand, the cosmological bound on $m_\nu$ may be somewhat
relaxed if new interactions of neutrinos exist, which led
to their enhanced annihilation in the early universe and diminished
the neutrino number density, or there is a new 
light long-lived particle whose 
electromagnetic decay might dilute the neutrino-to-photon ratio.
Moreover, a modification of the spectrum of the cosmological density
perturbations at a few Mpc scale would invalidate the most restrictive
cosmological bounds on $m_\nu$. So KATRIN would be helpful to diminish
the level of ambiguity.

The number of neutrino species is now best restricted by BBN,
$ {N_\nu = 3 \pm 0.3}$. The angular spectrum of CMBR is
less efficient, giving ${N_\nu = 3 \pm 1}$. But one can hope that
the Planck mission will be competitive with BBN or even will be
better than that. One should keep in mind, however, that the
number of the effective neutrino species
found from BBN and CMBR are sensitive 
to physics at different cosmological periods and they are not
not necessarily to coincide.

The heating of neutrinos by hotter electrons and positrons, when the
universe was about 1 sec old, distorted the spectrum 
of neutrinos and increased
their energy density by 3\%, plus 1\% from plasma corrections. If this
effect is observed in the future it will be the direct measurement of a 
physical process which took place in the very  early universe.

The bound on the cosmological lepton asymmetry for all neutrino flavors
found from BBN, was significantly improved when it was established that
the active neutrinos are strongly mixed. The maximum allowed value
for the leptonic chemical potential, $\mu/T < 0.07$,
makes it cosmologically insignificant.

The bounds on the mixing between active and possible sterile neutrinos
are significantly modified because of the strong mixing of all 
active neutrinos between themselves in comparison with earlier calculations,
when the mixing of active ones were not taken into account. 
Sometimes the new bounds are more restrictive than the old approximate ones
and in all cases are better than direct experiment results.

Both BBN and LSS formation are sensitive to violation of Fermi statistics
for neutrinos. If the bosonic fraction in neutrinos is larger than 
50\%, cosmic neutrinos may condence and make all dark matter in the
universe. Unfortunately there is no consistent theory of violation of 
the spin-statistics theorem and the consequences of its violation are not 
well justified.

Right handed neutrinos which must exist, if $m_\nu \neq 0$, are not allowed
to be abundant at BBN and one can derive from that quite strong bounds on the 
mass of possible right-handed intermediate bosons and mixing of them with the
left-handed ones. Similarly the bound on magnetic moment of neutrinos can be
derived. All these bounds are stronger than those found from direct experiment.
\\[3mm]
{\it Acknowledgement.} I thank Gary Steigman for critical comments.

\end{document}